\documentclass[twoside,a4paper,11pt]{sea10}
\usepackage{graphicx}
\usepackage{hyperref}
\usepackage{movie15}
\usepackage{natbib}
\def\teff{\mbox{$T_{\rm eff}$}}
\def\ebv{\mbox{$E(4405-5495)$}}
\def\rv{\mbox{$R_{5495}$}}
\def\al{\mbox{$A(\lambda)$}}
\def\mum1{\mbox{$\mu$m$^{-1}$}}

\begin{document}
\pagenumbering{arabic}
\pagestyle{myheadings}
\thispagestyle{empty}
{\flushleft\includegraphics[width=\textwidth,bb=58 650 590 680]{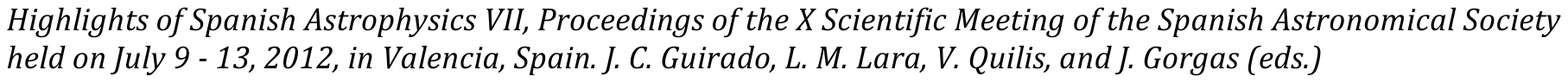}}
\vspace*{0.2cm}
\begin{flushleft}
{\bf {\LARGE
%
Do we need a new family of optical-NIR extinction laws?
%
}\\
\vspace*{1cm}
%
J. Ma\'{\i}z Apell\'aniz$^{1}$
%
}\\
\vspace*{0.5cm}
%
$^{1}$
Instituto de Astrof\'{\i}sica de Andaluc\'{\i}a-CSIC, Glorieta de la Astronom\'{i}a s/n, \linebreak 18008 Granada, Spain
%
\end{flushleft}
%
\markboth{
Do we need a new family of optical-NIR extinction laws?
}{ 
%
Ma\'{\i}z Apell\'aniz
%
}
\thispagestyle{empty}
\vspace*{0.4cm}
\begin{minipage}[l]{0.09\textwidth}
\ 
\end{minipage}
\begin{minipage}[r]{0.9\textwidth}
\vspace{1cm}
\section*{Abstract}{\small
%
I consider whether we can significantly improve the \citet{Cardetal89} family of extinction laws using new data and techniques. There are 
six different aspects that need to be treated: The use of monochromatic quantities, the three different wavelength regimes (NIR, optical and UV), the sample, 
and the photometric calibration. Excluding the behavior in the NIR and UV, I discuss the other four aspects and propose a new family of extinction laws 
derived from VLT/FLAMES and HST/WFC3 data.
%
\normalsize}
\end{minipage}
%
%

\section{The CCM family of extinction laws}

$\,\!$\indent The \citet{Cardetal89} or CCM family of extinction laws are the most oft-cited extinction laws in the astronomical literature.
Their fame is undoubtedly well deserved since they were the first laws that accurately described the full NIR-optical-UV range with a parameterized
form that allowed different tyes of extinction to be considered. Almost a quarter of a century later they are still broadly used in a wide range of 
applicatons. In this contribution we ask whether the CCM laws are due for an update using new data and techniques.

\begin{figure}
\centerline{
\includegraphics[width=0.83\linewidth, bb=28 50 566 530]{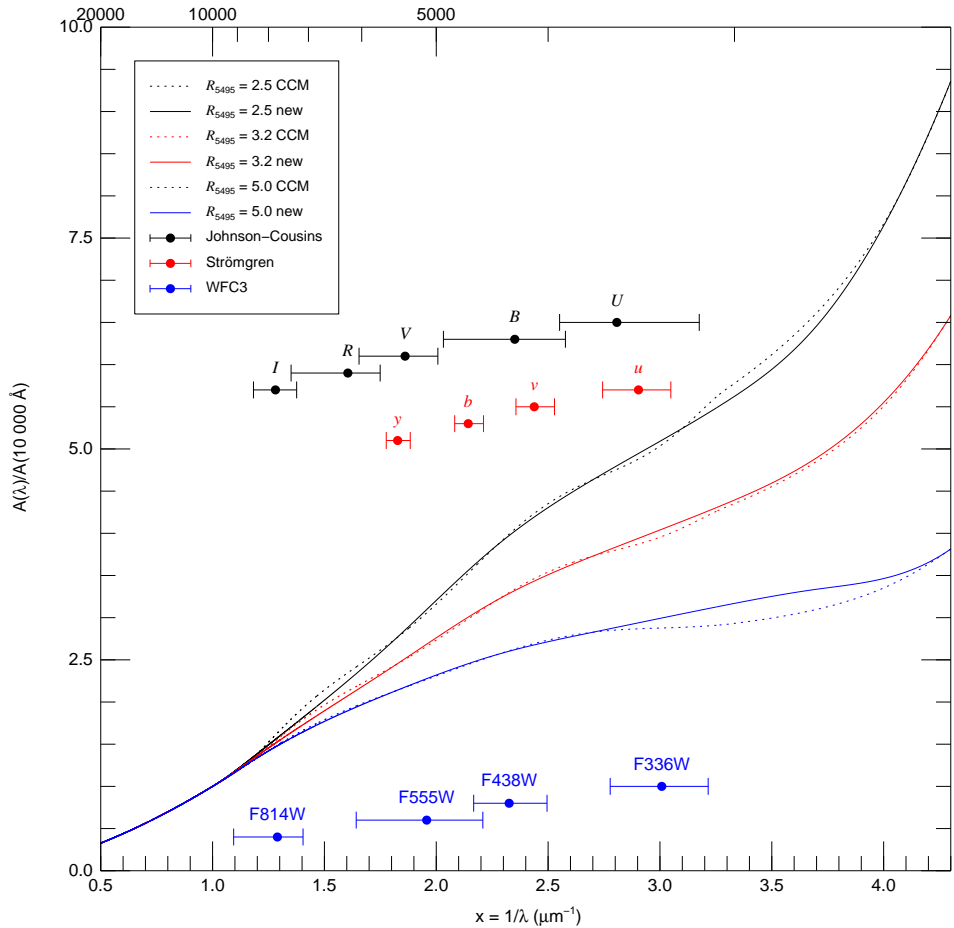}
}
\caption{CCM and new extinction laws for three values of \rv\ (2.5, 3.2, and 5.0). Extinction is normalized to the value at 10\,000 \AA\ in each 
case to emphasize that the extinction laws are the same for longer wavelengths and to better visualize the differences in the optical and NUV 
ranges. The approximate extent of some filters in three common systems (Johnson-Cousins, Str\"omgren, and WFC3) is shown.}
\label{extlaws1}
\end{figure}

The CCM extinction laws are a single-parameter family of functions that extend from $x\equiv 1/\lambda$ = 0.3 \mum1\ (33\,333 \AA) to 
$x$~=~10~\mum1\ (1000 \AA). The parameter that characterizes the family is $R_V\equiv A_V/E(B-V)$, which has typical values close to 3 but which
in some environments can be slightly lower or significantly higher. The CCM extinction laws are divided into four wavelength ranges, the NIR 
($x$ = 0.3-1.1 \mum1), the optical ($x$ = 1.1-3.3 \mum1), the UV ($x$ = 3.3-8.0 \mum1), and the FUV ($x$~=~8.0-10.0~\mum1). Each range uses a
different functional form but the laws are continuous and differentiable at each boundary. The extinction laws in the NIR and optical were derived
from multiband ground-based photometry and in the UV and FUV from IUE spectrophotometry. Three examples of CCM extinction laws are shown in
Figs.~\ref{extlaws1}~and~\ref{extlaws2}.

There are several issues with the CCM laws that need to be discussed in other to study the need for a new family of extinction laws:

\begin{figure}
\centerline{
\includegraphics[width=0.89\linewidth, bb=28 50 566 530]{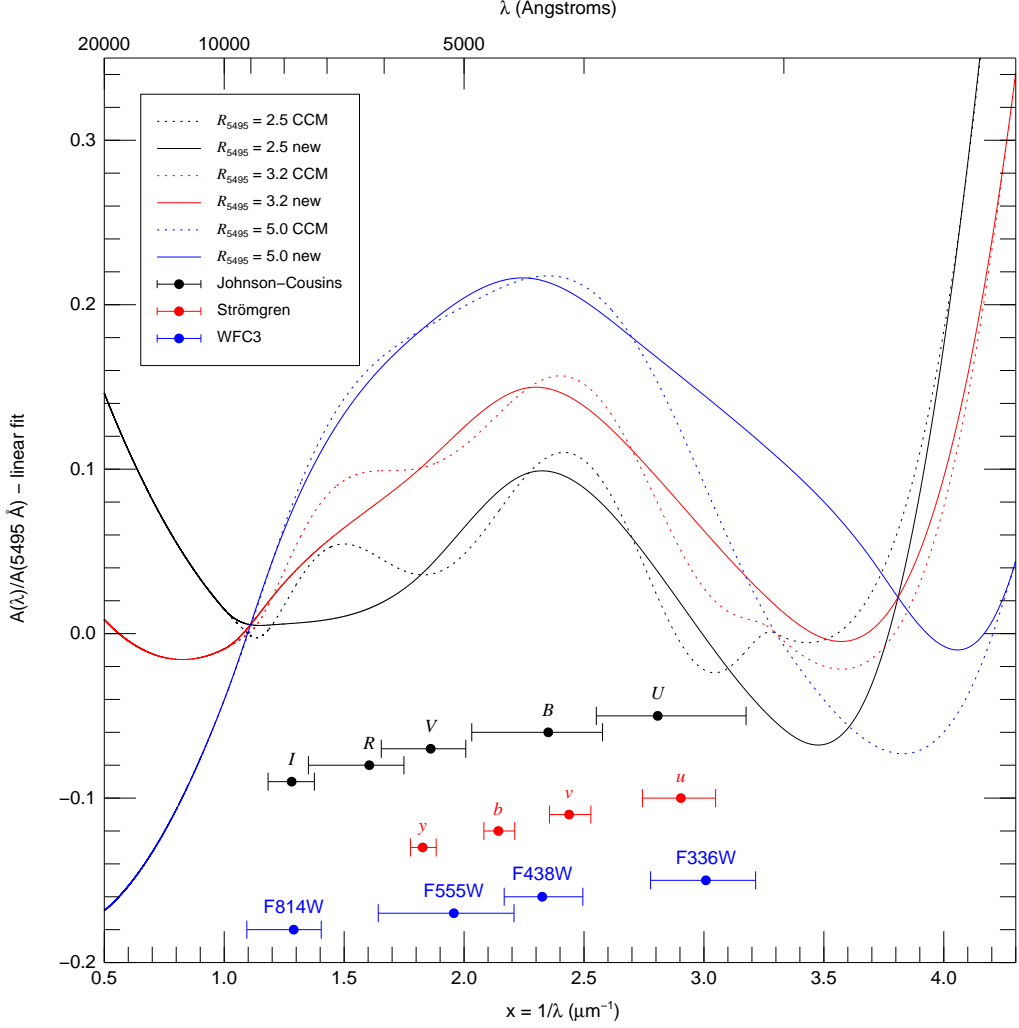}
}
\caption{As Fig.~\ref{extlaws1} but with a different normalization to emphasize the differences between extinction laws. In each case we have 
subtracted a linear fit $A(\lambda)/A(5495 \AA) = a(\rv) + b(\rv)x$, with $a(\rv)$ and $b(\rv)$ calculated so that the CCM law for that \rv\ is 0.0 
at $x=1.1$ and at $x=3.3$, the limits for the optical range in CCM.}
\label{extlaws2}
\end{figure}

\begin{figure}
\centerline{
\includegraphics[width=0.550\linewidth, bb=28 35 556 566]{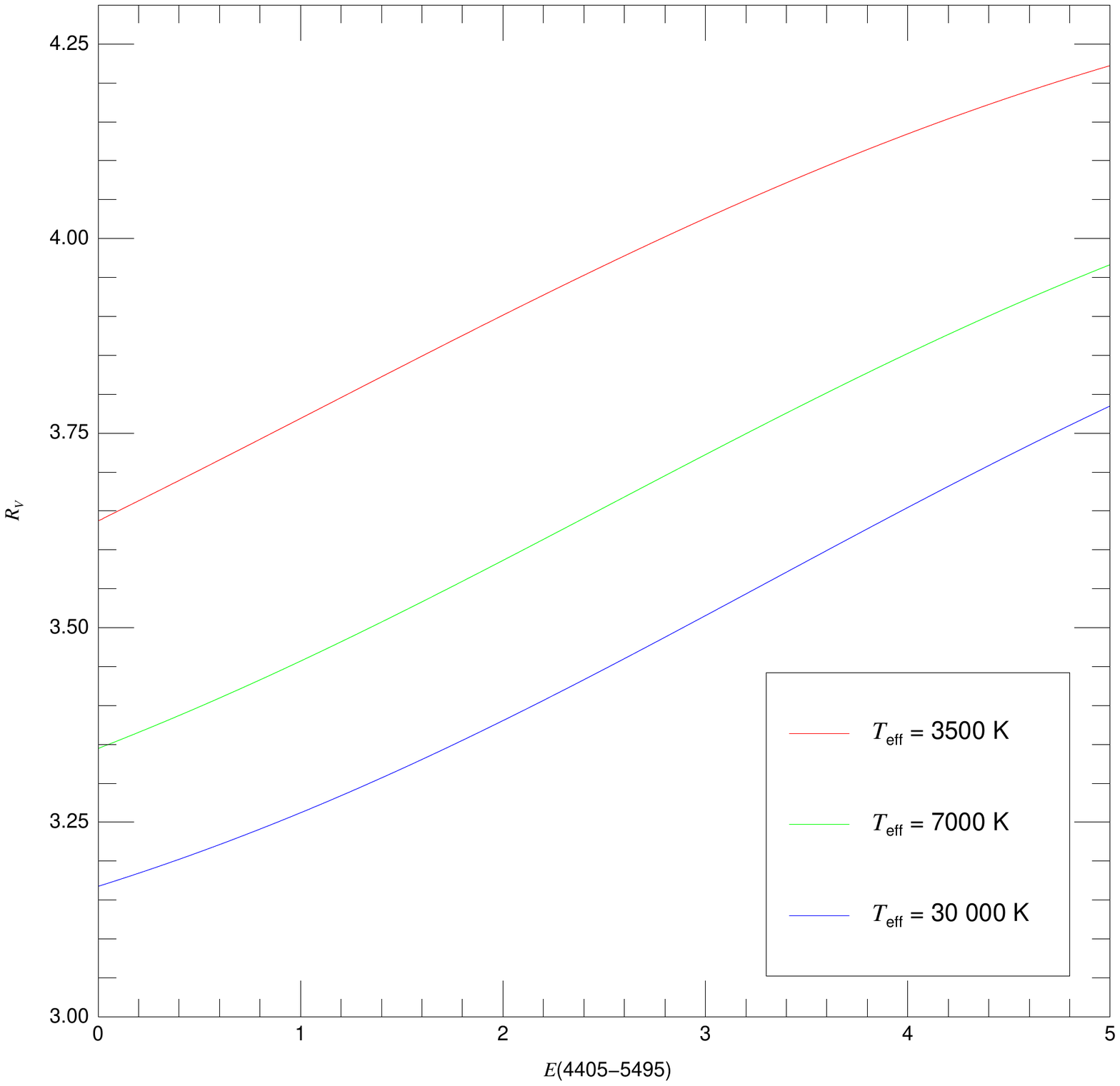} ~ 
\includegraphics[width=0.558\linewidth, bb=38 40 566 566]{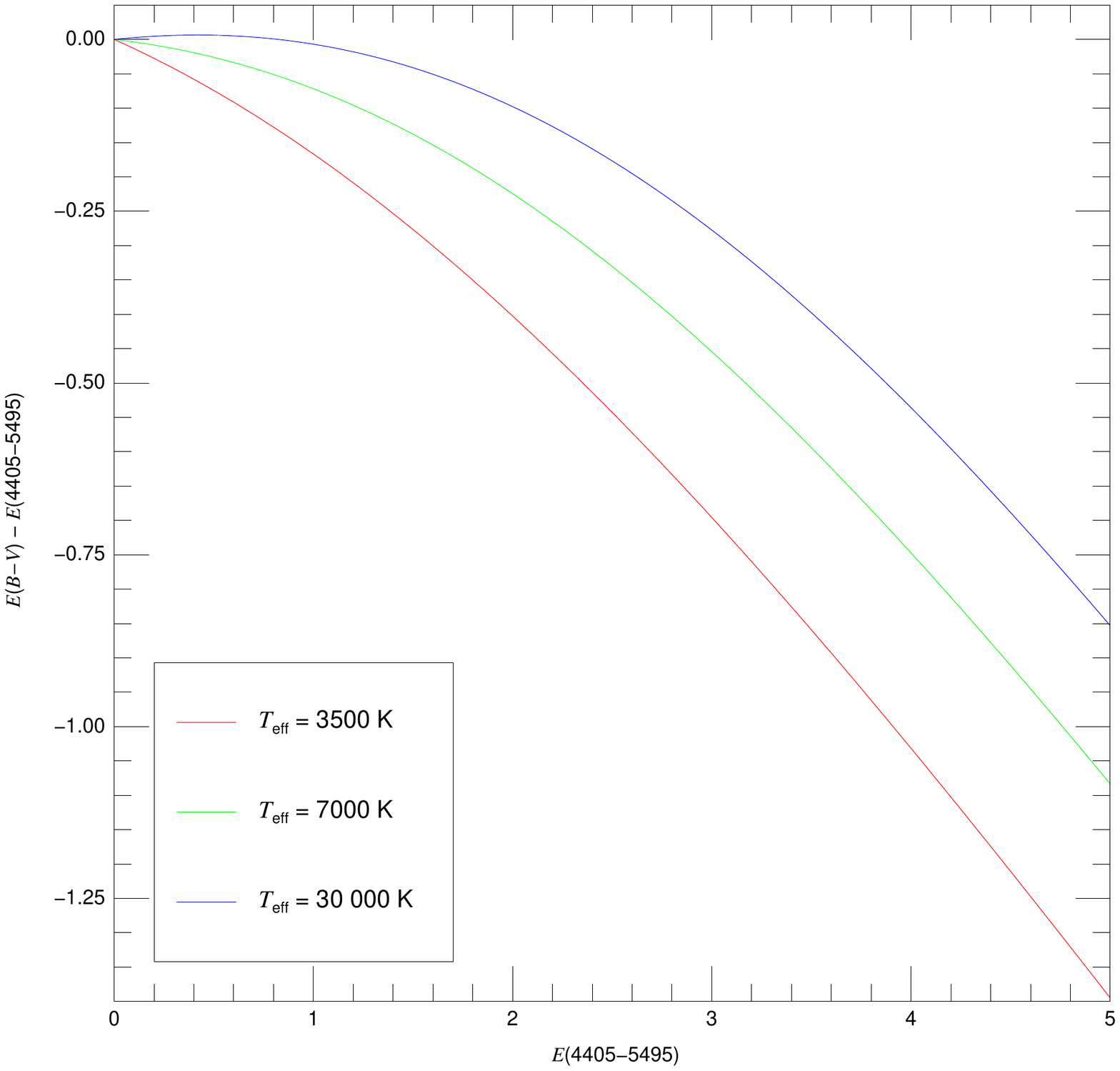}
}
\caption{$R_V$ (left) and $E(B-V) - \ebv$ (right) as a function of \ebv\ for a CCM extinction law with \rv\ = 3.2 and three main-sequence stars with different \teff.
$R_V \approx \rv$ and $E(B-V) \approx \ebv$ only for hot stars with low extinctions.}
\label{rvebv}
\end{figure}

	{\bf 1. Monochromatic quantities.} Using $R_V$ as the parameter for the type of extinction is ill defined. $R_V$ is a filter-integrated 
quantity (using Johnson $B$ and $V$), not a monochromatic one\footnote{A quantity will be called ``monochromatic'' when it is defined based only on
discrete wavelengths.}. That 
means that unless \al\ is constant over the extent of the filter (which is never the case), $R_V$ is not only a function of the extinction law but 
also of the input spectral energy distribution and of the amount of extinction. In other words, the same amount and type of dust in front of two 
stars produces different values of $R_V$ and doubling the amount of dust without changing its type also changes $R_V$. For the same reason, using 
$A_V$ or $E(B-V)$ to parameterize the amount of extinction is also ill defined. Graphical examples of the problem are shown in 
Fig.~\ref{rvebv}.

	So what is going on here? Actually, a CCM law with a given $R_V$ parameter does not actually produce extinction with $A_V/(E(B-V)$ equal to 
that value of $R_V$ (see Fig.~\ref{rvebv}), an effect that has created confusion in the past. The easiest way to fix it is to select two wavelengths 
near the center of the $B$ and $V$ filter and substitute the $R_V$ (type of extinction) and $E(B-V)$ (amount of extinction) parameters by
monochromatic equivalents. \citet{Maiz04c} did this choosing 4405 \AA\ and 5495~\AA as reference wavelengths for $B$ and $V$, respectively. 
Those are also the choices in the current version of CHORIZOS (v 3.2, \citealt{Maiz04c}) 
and here we follow the same convention. The reason for using those wavelengths to define the equivalents of $R_V$ and $E(B-V)$,
e.g. \rv\ and \ebv, is that for hot stars and low extinction [\ebv $<$ 1.0], $R_V \approx \rv$ and $E(B-V) \approx \ebv$ 
(Fig.~\ref{rvebv}). Therefore, we will use \rv\ as the parameter for both the CCM and new laws.

	{\bf 2. The NIR.} In that wavelength range (called simply infrared in the original paper), CCM laws use a power law with a 
fixed\footnote{This is not obvious in Fig.~3 of CCM because the extinction law there is normalized with respect to $V$ or, more properly, 5495 \AA.}
exponent $\al/A(10\,000\,\AA) = x^{1.61}$. However, their analysis has been put into question by \citet{Nishetal09} and different authors 
(e.g. \citealt{Mooretal05,FitzMass09}) have found that the NIR extinction law changes from one sightline to another, with exponents as low as 1.1
and as high as 2.3. Beyond the $K$ band, the extinction law first flattens and then develops complex structures 
\citep{Romaetal07,Nishetal09,Gaoetal09}.

	{\bf 3. The optical range.} In order to fit the extinction law there, CCM used a seventh-degree polynomial. That functional form has the
advantage of being able to fit the law through their five passbands ($UBVRI$) but the disadvantage of possibly introducing undesired wiggles in
wavelength (Fig.~\ref{extlaws2}). Such wiggles can be detected when comparing CCM predictions with spectrophotometry or with intermediate-band
photometry such as Str\"omgren's. Prior to CCM, \citet{Whit58} (see also \citealt{ArdeVird82}) had used spectrophotometry to propose a simpler 
functional form of two straight lines joined near $x$~=~2.25~\mum1. The point where the two lines are joined is called the ``knee'' and is the most 
prominent wiggle in the CCM laws in the optical (Fig.~\ref{extlaws2}, see also \citealt{StebWhit43}). CCM were clearly aware of the issue and mentioned 
that ``The Whitford law may therefore be more accurate for the diffuse ISM near $x \approx$ 2.25 \mum1. The virtue of ours, however, is that 
it joins smoothly onto the UV extinction law from the FM [\citealt{FitzMass88}] sample of stars and that it takes into account the {\it differences 
in the extinction laws of lines of sight with various values of $R_V$}''. In other words, they were willing to sacrifice accuracy (in the form of
detailed behavior for small wavelength scales) in favor of functionality (in the form of the addition of the $R_V$ parameter and the continuity and
differentiability for all wavelengths). 

	{\bf 4. The UV and FUV.} The emphasis of CCM laid on these wavelength ranges, as appropriate at the end of a decade where IUE revolutionized
the study of extinction. Special importance was paid to the ability of a single-parameter family to describe extinction from the IR to UV, as evident
in the title of the paper. However, this aspect has been challenged by \citet{FitzMass07}, who find that ``With the exception of a few curves with 
extreme (i.e., large) values of $R(V)$, the UV and IR portions of Galactic extinction curves are not correlated with each other.''

{\bf 5. The CCM sample.} CCM used a sample of 29 stars, which was a 
relatively large number for the time. However, more modern extinction studies use hundreds or thousands of stars. Also, their sample had only low or
moderate extinction: only three stars have values of $(E(B-V)$ greater than 1.0 and even the most reddened (HDE 229\,196) of those has an $E(B-V)$ 
of 1.22. Also, the three most reddened stars have a relatively narrow range of \rv\ between 2.6 and 4.2. The solution for high values of \rv\ ($>$
5.0) depends heavily on a single star, Herschel 36, which has an $E(B-V)$ of 0.89 (the other cases with \rv\ $>$ 5.0 have $E(B-V) <$ 0.4) and has
been recently found to be a multiple system with nearby IR sources \citep{Ariaetal06,Ariaetal10}. Using stars with low extinction to determine an
extinction law is quite dangerous since random and systematic (see below) uncertainties can introduce significant errors. One should also take into
account that the CCM law has been applied to objects with \ebv\ significantly larger than 1.0: in those cases one is extrapolating a law beyond the
range for which it was derived, thus amplifying any potential errors in the original work.

	{\bf 6. Photometric calibration.} The final issue is that of the photometric calibration of the filters used in the NIR and optical ranges.
By calibration we mean both the determination of the zero points and the shape of the sensitivity curves. The reader is referred to
\cite{Maiz05b,Maiz06a,Maiz07a} and references therein for developments in this field after CCM. In particular, it should be noted that old
observations in the Johnson $U$ band may be suspect because of the effect of the atmosphere in the sensitivity curve and because the curve straddles
the Balmer jump, thus being affected by errors in the spectral type.

\section{The new extinction laws}

$\,\!$\indent 	Based on the issues discussed in the previous section, we decided to attempt the calculation of a new
family of extinction laws. Ideally, to complete such a task one would use high-quality spectrophotometry from the NIR to the UV of a diverse
collection of sources in different environments and with different degrees of extinction. Since such dataset is not currently available, we will start
by dealing only with some of the issues discussed in the previous section. More specifically, we will ignore extinction in the UV (except for the
region closest to the optical) and for the NIR we will simply use the CCM laws. In other words, we will concentrate our efforts in the optical region.
Ignoring the UV will not matter to a non-specialist interested only in
eliminating the extinction from his/her optical data. Ignoring the NIR may matter if the exponent there is significantly different from the CCM one
but only if extinction is very large and even then it may only apply to the total extinction correction, not to e.g. the determination of \teff\ from
the photometry.

To derive the new family of extinction laws we are using two datasets:

\begin{itemize} 
 \item The VLT/FLAMES Tarantula Survey \citep{Evanetal11a}, from which we are using $\sim 200$ accurate spectral types of stars (mostly of O type) 
       in 30 Doradus as well as NIR photometry.
 \item The WFC3 Early Release Science HST/WFC3 images of the central region of 30 Doradus \citep{DeMaetal11a,Sabbetal12}, from which photometry in six broadband
       filters (F336W, F438W, F555W, F814W, F110W, and F160W, which are $UBVIJH$ equivalents) was extracted.
\end{itemize}

Our analysis is in a quite advanced stage and in the near future we will submit for publication our results. A preliminary view with
three examples of the new extinction laws is shown in Figs.~\ref{extlaws1}~and~\ref{extlaws2}. We have verified that the new laws provide significantly lower residuals
when assuming effective temperatures derived from the spectral types and better fits when attempting to derive the effective temperatures from the photometry. We have
also applied the new extinction laws to Galactic stars with \ebv\ between 1.5 and 2.0 and have been able to significantly lower the fits residuals when compared to CCM
or to \citet{Fitz99}, thus extending the applicability of the new family of optical/NIR extinction laws to the Milky Way\footnote{Note that the extensive literature on
the differences between the SMC, LMC, and MW extinction laws refers to the UV regime, not to the optical/NIR.}

%
%
\small  
%
\section*{Acknowledgments}   
%

This work is based on observations at the European Southern Observatory Very Large Telescope in programme 182.D-0222 and on
observations made with the NASA/ESA Hubble Space Telescope (HST) associated with GO program 11360 and obtained at the Space 
Telescope Science Institute, which is operated by the Association of Universities for Research in Astronomy, Inc., under NASA 
contract NAS 5-26555. I would like to thank Max Mutchler for helping with the processing of the WFC3 images. I acknowledge
support from [a] the Spanish Government Ministerio de Educaci\'on y Ciencia through grants AYA2010-15081 and AYA2010-17631,
[b] the Consejer{\'\i}a de Educaci{\'o}n of the Junta de Andaluc{\'\i}a through grant P08-TIC-4075, and [c] the George P. and 
Cynthia Woods Mitchell Institute for Fundamental Physics and Astronomy. I am also grateful to the Department of Physics and 
Astronomy at Texas A\&M University for their hospitality during some of the time this work was carried out. Finally, I thank Chris
Evans, the rest of the VLT/FLAMES Tarantula Survey consortium, and Karl Gordon for our discussions on this topic.

%
%
%
%
%
\bibliographystyle{aj}
\small
\bibliography{general}

\end{document}